\documentclass[pre,10pt]{revtex4}
\usepackage{amsmath}    % need for subequations
\usepackage{amsfonts}
\usepackage{amssymb}	
\usepackage{graphicx}   % need for figures
\usepackage{verbatim}   % useful for program listings
\usepackage{color}      % use if color is used in text
\usepackage{subfigure}  % use for side-by-side figures
\usepackage{hyperref}   % use for hypertext links, including those to external documents and URLs

\usepackage{amsmath}

\hypersetup{pdfborder = {0 0 0},colorlinks=true}
\newcommand{\upd}{\mathrm{d}}
\begin{document}

\title{A statistical mechanics approach to the sample deconvolution problem}
%\shorttitle{Deconvolution of gene expression data} %Insert here a short version of the title if it exceeds 70 characters

\author{N. Riedel and J. Berg}
\email{nriedel@thp.uni-koeln.de and berg@thp.uni-koeln.de}

\affiliation{                    
  Institut f\"ur Theoretische Physik, University of Cologne - Z\"ulpicher Stra\ss e 77, 50937 K\"oln, Germany\\
Sybacol, University of Cologne, Germany
}

\begin{abstract}
In a multicellular organism different cell types express a gene in different amounts. Samples from which gene expression levels can be measured typically contain a mixture of different cell types, the resulting measurements thus give only averages over the different cell types present. Based on fluctuations in the mixture proportions from sample to sample it is in principle possible to reconstruct the underlying expression levels of each cell type: to deconvolute the sample. We use a statistical mechanics approach to the problem of deconvoluting such partial concentrations from mixed samples, give analytical results for when and how well samples can be unmixed, and suggest an algorithm for sample deconvolution.
\end{abstract}

\maketitle

\section{Introduction}
Organs in higher organisms are complex tissues containing a variety of different cell types. Brain tissue, for instance, contains not only neurons, but also supporting cells as the astrocytes and oligodendrocytes.  Kidney tissue contains the filtering units (podocytes) as well as cells of the capillary system (tubules). Whereas two cells of different cell type have largely the same DNA sequence, only a cell-type specific set of genes will be expressed in a cell~\cite{b.Palmer2006,b.Galbraith2006}.

Over the last two decades, experimental methods have been developed which allow to measure the amount of m(essenger)-RNA from different genes in a sample~\cite{b.Schena1995,b.Lashkari1997}. However, one may not be interested in expression levels averaged over all cell types in such a sample, but may want to know the mRNA levels present in the different cell types. A particularly pressing example arises in cancer research, where tissue samples typically contain solid tumour and healthy tissue in unknown proportions~\cite{s.Cleator2006}. 

Denoting the proportion of cell type $a$ in sample $\mu$ by $p_a^\mu$, and the concentration of mRNA from gene $i$ in cell type $a$ by $x_i^a$, the concentration of mRNA from gene $i$ in sample $\mu$, $X^\mu_i$ is given by
\begin{equation}
\label{eq.samplecoupling1}
 X^\mu_i=\sum_{a=1}^n  p_a^\mu x_i^a + \xi_i^\mu \ ,
\end{equation}
where the so-called residuals $\xi_i^\mu$ stem from sample-specific fluctuations of concentrations or random experimental errors. The number of different cell types is denoted by $n$. Additionally, we have the constraints $0<x_i^a$, $0<p_a^\mu<1$ and $\sum_a p_a^\mu = 1 \,\forall\,\mu$. 

\begin{figure}
\includegraphics[width=80.0mm]{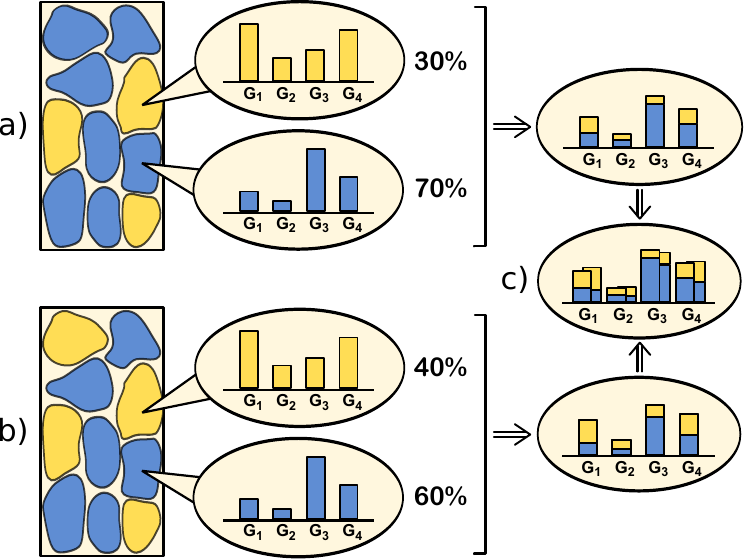}
\caption{{\bf The sample deconvolution problem.} a) A tissue sample containing a mixture of different cell types is taken. The concentration of a particular gene product in the sample is a linear combination of the concentrations in each cell type. b) Two different tissue samples generally  contain different mixtures of cell types, while the concentration of gene products is largely constant  across cells of a given type. c) Part of the variation of expression levels across samples is 
due to fluctuations in the mixing proportions. This provides the basis for reconstructing concentrations in each cell type and mixing proportions.}
\label{fig.ConvolutionExample}
\end{figure}

Sample deconvolution~\cite{j.Clarke2010} is the inverse problem of reconstructing the concentrations of gene products $x_i^a$ in each cell type, as well as the mixing proportions $p_a^\mu$ of the samples from measurements of mixed samples $X^\mu_i$. The information necessary for this reconstruction must come from fluctuations in the mixing proportions across samples: A cell type with high concentrations of mRNA of genes $i$ and $j$ will induce positively correlated fluctuations of the measurements $X^\mu_i$ and $X^\mu_j$ as the fraction of this cell type varies from sample to sample (see fig.~\ref{fig.ConvolutionExample}). 

Equation~(\ref{eq.samplecoupling1}) also arises in a broad range of contexts outside molecular biology. It is the fundamental equation of \textit{factor analysis}, a statistical approach where different unknown factors $x_a$ contribute with linear weights $p_a$ to some outcome $X$~\cite{d.Barber2012}. Applications arise for example in the context of face recognition~\cite{s.Prince2007}, data analysis in ecology~\cite{a.Hirzel2002}, or fluorescence microscopy~\cite{b.Zimmermann2005,b.Lansford2001}.

There are two questions we address in this paper. First, under what conditions is such learning from fluctuations possible at all? And second, how accurately can the reconstruction be made? We first discuss a general constraint on the minimal number of samples needed from linear algebra. We then formulate a Bayesian model for 
sample deconvolution and study this model from the point of view of statistical physics. We derive analytical results for the accuracy of reconstruction for a simple, non-trivial case of the problem. This model also suggests a simple general algorithm based on Markov chain Monte Carlo (MCMC) sampling. Comparison with an established class of algorithms~\cite{b.Repsilber2010,b.Lahdesmaki2005,b.Venet2001}, which formulate sample deconvolution as an optimisation problem, demonstrates a generic drawback of such optimization approaches.

As an initial remark, we derive a constraint on the minimal number of samples needed for reconstruction from linear algebra. Looking at the number of variables in eq.~(\ref{eq.samplecoupling1}) we have, neglecting the residuals, the matrix equation
\begin{alignat}{3}
\label{eq.VariableCounting}
&\underbrace{\begin{pmatrix}  X_1^1 & \cdots & &\\ \vdots & \ddots & &\\  & & &\\  & & &\\ \end{pmatrix}}
&=\underbrace{\begin{pmatrix}  p_1^1 & \cdots \\ \vdots & \ddots\\  & \\  & \\\end{pmatrix}}
&\times && \underbrace{\begin{pmatrix}  x_1^1 & \cdots & &\\ \vdots & \ddots & &\\ \end{pmatrix} },\nonumber\\
&\quad\qquad MN             &\geq    M(n-1)\quad &+  && \quad\qquad nN  \ .
\end{alignat}
Denoting the total number of samples by $M$, the number of genes by $N$ and the number of cell types by $n$, there are $MN$ measurements on the left and $M(n-1)+nN$ unknown variables on the right-hand side. If the number of unknown variables exceeds the number of data points, the system of equations is underdetermined. For a measurement of gene expression levels, the number $N$ of genes will typically be of the order of hundreds or even thousands, exceeding by far the number $n$ of cell types in a sample, or the number $M$ of samples. For $N \gg (n,M)$ the condition not to have an underdetermined set of equations reduces to $M > n$, so the number of samples taken has to be at least larger than the number of cell types.

\section{Bayesian model for sample deconvolution}

For concreteness we assume that the residuals $\xi_i^\mu$ in (\ref{eq.samplecoupling1}) are independent and identically distributed Gaussian variables. 
Other distributions will be discussed below. These residuals stem from fluctuations in the concentrations of the same cell type (`biological noise') and random experimental error (`technical noise'). Given the mixing proportions $p_a^\mu$ and the concentrations in cell types $x_i^a$, this distribution of residuals induces the distribution of the measurements $X^\mu_i$  
\begin{equation}
\label{eq:likelihood}
P\left( \mathbf{X} \vert \mathbf{p}, \mathbf{x}\right ) = \left(2\pi \sigma_\xi^2\right)^{-\frac{MN}{2}} e^{-\mathcal{H}\left(\mathbf{p},\mathbf{x};\mathbf{X}\right)}\,,
\end{equation}
with the Hamiltonian 
\begin{equation}
\label{eq.Hlikelihood}
\mathcal{H}\left(\mathbf{p},\mathbf{x};\mathbf{X}\right) = \sum\limits_{\mu, i} \frac{\left( X^\mu_i - \sum_a p_a^\mu x_i^a\right)^2}{2\sigma_\xi^2} \,.
\end{equation}
Bold symbols are used to denote the corresponding matrices. The quantity of interest, namely the probability of a given set of mixtures and of  concentrations in cell types given the measurements, 
follows from Bayes theorem; the so-called posterior probability $P\left( \mathbf{p}, \mathbf{x} \vert \mathbf{X} \right ) =  \frac{P \left( \mathbf{p}, \mathbf{x} \right ) P\left( \mathbf{X} \vert \mathbf{p}, \mathbf{x}\right )}{P\left( \mathbf{X} \right )}$ is expressed in terms of the likelihood (\ref{eq:likelihood}), the prior $P \left( \mathbf{p}, \mathbf{x} \right )$ and the marginal likelihood $P\left( \mathbf{X} \right )$ (which for a given set of measurements is just a multiplicative constant). 

For the prior $P \left( \mathbf{p}, \mathbf{x} \right )$ a particular choice has to be made. To keep the analytical calculations tractable, we choose again independent Gaussian distributions with means $\bar{x}_a$/$\bar{p}_a$ and variances $\sigma_{x,a}^2$/$\sigma_{p,a}^2$, where we allow for different distribution parameters for each cell type. Here, we assume that the distribution parameters are chosen such that the contributions of values $x_i^a<0$ and $p_a^\mu<0$ or $p_a^\mu>1$ are negligibly small. In the general algorithm we discuss below, any distribution can be implemented. The constraint that mixing proportions for each sample add up to one is implemented using a delta function. 
For this particular choice of the prior we get
\begin{equation}
\label{eq.Posterior}
P\left( \mathbf{p}, \mathbf{x} \vert \mathbf{X} \right ) =  \left[ \prod_\mu \delta\left(\sum_{a=1}^n p_a^\mu-1\right)\right]  e^{-\mathcal{H}_{\rm{Bayes}}\left(\mathbf{p},\mathbf{x};\mathbf{X}\right)} /Z_\mathbf{X} \,,
\end{equation}
with the Hamiltonian
\begin{align}
\label{eq.Hprior}
 &\mathcal{H}_{\rm{Bayes}}\left(\mathbf{p},\mathbf{x};\mathbf{X}\right) = \sum_{\mu i} \frac{\left(X^\mu_i - \sum_{a} p_a^\mu x_i^a\right)^2}{2\sigma_\xi^2}\nonumber\\ 
&+ \sum_{\mu a} \frac{\left(p_a^\mu-\bar{p}_a \right)^2}{2\sigma_{p,a}^2} + \sum_{a i} \frac{\left(x_i^a-\bar{x}_a \right)^2}{2\sigma_{x,a}^2}. 
\end{align}
The first term in eq.~(\ref{eq.Hprior}) comes from the likelihood, penalizing the deviation of a possible solution $\{p_a^\mu, x_i^a\}$ from the measurement $\{X^\mu_i\}$. The second and third term are the Gaussian priors for the mixing proportions and the expression patterns, respectively. The posterior distribution~\eqref{eq.Posterior} describes the state of our knowledge of the mixing proportions and concentrations in each cell type, given the measurements. From the perspective of statistical physics, the  mixing proportions and concentrations in each cell type define a phase space, and the posterior distribution~\eqref{eq.Posterior} defines a Hamiltonian~\eqref{eq.Hprior} describing how strongly the probability measure of mixing proportions/concentrations is focused in particular parts of this phase space. The partition function 
\begin{equation}
\label{eq:partition}
Z_\mathbf{X}(\beta) = P\left( \mathbf{X} \right ) = \mathrm{Tr}_{\mathbf{p},\mathbf{x}}\, e^{-\beta \mathcal{H}_{\rm{Bayes}}(\mathbf{p},\mathbf{x};\mathbf{X})},
\end{equation}
with $\mathrm{Tr}_{\mathbf{p},\mathbf{x}} = \int \upd \mathbf{p} \int \upd  \mathbf{x} \prod_\mu \delta\left(\sum_{a=1}^n p_a^\mu-1\right)$ sets out the statistical mechanics of sample deconvolution at $\beta=1$. The corresponding entropy $S=\left. \frac{\partial}{\partial \beta} \frac{1}{\beta} \ln Z_\mathbf{X}\right|_{\beta=1}$ is a measure of the uncertainty of reconstruction.

\section{Partition function and theoretical reconstruction accuracy}

We can now address the theoretical limit of sample deconvolution. For a given set of measurements $\mathbf{X}$, the statistics of mixtures and concentrations is described by the partition function~\eqref{eq:partition}. Suppose now that these measurements are generated using~\eqref{eq:likelihood} from underlying mixing proportions $\mathbf{p}^T$ and concentrations $\mathbf{x}^T$. These are the targets the reconstruction aims for. In order to explore the behaviour of the system for typical realisations of these targets, the quenched average of $\ln Z_\mathbf{X}$ over $\mathbf{p}^T$ and $\mathbf{x}^T$ needs to be computed~\cite{b.Mezard1987}. We restrict ourselves to the so-called annealed approximation and calculate the average of $Z_\mathbf{X}$ over the target mixtures and concentrations. We will show numerically that the difference between quenched and annealed results are small for a reasonable choice of parameters. The annealed average over $Z_\mathbf{X}$ is given by
\begin{align}
&\left\langle\langle Z_\mathbf{X} \right\rangle\rangle = \int \upd \mathbf{p}^T \int \upd \mathbf{x}^T \int \upd \boldsymbol{\xi} \int \upd \mathbf{X}\, P(\mathbf{p}^T,\mathbf{x}^T, \boldsymbol\xi) \nonumber\\
&\left[\prod_\mu  \delta\left(\sum_{a=1}^n p^T_{\mu a}-1\right)\right] \delta\left(\mathbf{X}-\mathbf{p}^T\mathbf{x}^T-\boldsymbol\xi\right)\,\,Z_\mathbf{X}\, ,
\label{eq.PartitionAverage}
\end{align}
The average $\langle\langle . \rangle\rangle$ is over all data matrices weighted by their probabilities under the generative model $P(\mathbf{p}^T,\mathbf{x}^T, \boldsymbol\xi)=\exp\left\lbrace - \frac{1}{2\sigma_\xi^2}  \boldsymbol\xi^2 - \frac{1}{2\sigma_x^2} \left( \mathbf{x^T} - \bar{\mathbf{x}} \right)^2 - \frac{1}{2\sigma_p^2} \left( \mathbf{p^T} - \bar{\mathbf{p}} \right)^2 \right\rbrace$ for given parameters $\bar{x}_a$/$\bar{p}_a$, $\sigma_{x,a}^2$/$\sigma_{p,a}^2$ and $\sigma_\xi^2$. This average of the partition function $Z_\mathbf{X}$ leads to a large set of Gaussian integrals, which in the thermodynamic limit $N\rightarrow\infty$ can be evaluated using the saddle-point approximation~\cite{b.Mezard1987,b.Gardner1988_1,b.Gardner1988_2,m.Opper1995}. For the thermodynamic limit we consider the scaling ansatz $M=\alpha N$ and $\sigma_\xi^2=\tilde{\sigma}_\xi^2 N$. In a concrete situation $N$ is of course finite, and $\alpha, \tilde{\sigma}_\xi$ will be small (but also finite). In a lengthy but standard calculation we obtain the saddle point equations
\begin{alignat}{2}
\label{eqn:saddlepointeqns}
\hat{q}_{a b} &= \frac{\alpha}{4\tilde{\sigma}_\xi^2} \left\langle p_a p_b \right\rangle_{\mathbf{p},\mathbf{p}^T},  \quad &q_{a b} &= \left\langle x_a x_b \right\rangle_{\mathbf{x},\mathbf{x}^T}, \nonumber \\[0.2cm] 
\hat{q}_{a b}^T &= -\frac{\alpha}{2\tilde{\sigma}_\xi^2} \left\langle p_a^T p_b \right\rangle_{\mathbf{p},\mathbf{p}^T}, \quad &q_{a b}^T &= \left\langle x_a^T x_b \right\rangle_{\mathbf{x},\mathbf{x}^T}, \\[0.2cm] 
\hat{q}_{a b}^{TT} &= \frac{\alpha}{2\tilde{\sigma}_\xi^2} \left\langle p_a^T p_b^T \right\rangle_{\mathbf{p},\mathbf{p}^T}, \quad &q_{a b}^{TT} &= \left\langle x_a^T x_b^T \right\rangle_{\mathbf{x},\mathbf{x}^T}, \nonumber
\end{alignat}
with averages defined as
\begin{align}
&\left\langle \left(\cdots\right) \right\rangle_{\mathbf{p},\mathbf{p}^T} =  \frac{1}{Z_p} \int \prod_a d\,p_a^T \int \prod_a d\,p_a \,\, \left(\cdots\right) \nonumber \\
&\delta\left(\sum_{a=1}^n p_{a}^T-1\right) \delta\left(\sum_{a=1}^n p_{a}-1\right) \nonumber \\
&\exp\Bigg\lbrace - \frac{1}{4\tilde{\sigma}_\xi^2} \sum_{a b} \left( p_a p_b q_{a b} + p_a^T p_b^T q_{a b}^{TT} -2 p_a^T p_b q_{a b}^T \right) \nonumber \\
&- \frac{1}{2\sigma_p^2} \sum_a \left(p_a-\bar{p}\right)^2 - \frac{1}{2\sigma_p^2} \sum_a \left(p_a^T-\bar{p}\right)^2 \Bigg\rbrace
\end{align}
and
\begin{align}
&\left\langle \left(\cdots\right) \right\rangle_{\mathbf{x},\mathbf{x}^T} = \frac{1}{Z_x} \int \prod_{a}d\,x^T_{a} \int \prod_{a}d\,x_{a}
 \left(\cdots\right) \nonumber \\
&\exp\Bigg\lbrace - \sum_{a b} \left( x_a x_b \hat{q}_{a b} + x^T_a x_b \hat{q}^T_{a b} + x^T_a x^T_b \hat{q}^{TT}_{a b} \right) \nonumber \\
&- \frac{1}{2\sigma_x^2} \sum_{a}\left( x_{a} - \bar{x}\right)^2 - \frac{1}{2\sigma_x^2} \sum_{a}\left( x^T_{a} - \bar{x}\right)^2 \Bigg\rbrace\,.
\end{align}
The $q$-variables in eq. (\ref{eqn:saddlepointeqns}) are the order parameters of the system. They are connected to the Euclidean distance between the reconstruction and its target, $r_x = \frac{1}{nN} \sum_{a,i} \left(x_i^a-x_i^{T a}\right)^2$ and similarly $r_p = \frac{1}{nM} \sum_{a,\mu} \left(p_a^\mu-p_a^{T \mu}\right)^2$, through the relationships $r_x=\frac{N}{n}\sum_a \left( q_{aa} - 2 q_{aa}^T + q_{aa}^{TT} \right)$ and $r_p=\frac{N}{n}\sum_a \left( \hat{q}_{aa} - 2 \hat{q}_{aa}^T + \hat{q}_{aa}^{TT} \right)$. These saddle-point equations have to be solved numerically. We note the formal similarity of the number of cell types with replicas used to calculate quenched averages. A quenched calculation of this system would thus result in a system of equations bearing a two-replica structure \cite{r.Monasson1995}. 

\begin{figure}
\begin{center}
\includegraphics[width=80.0mm]{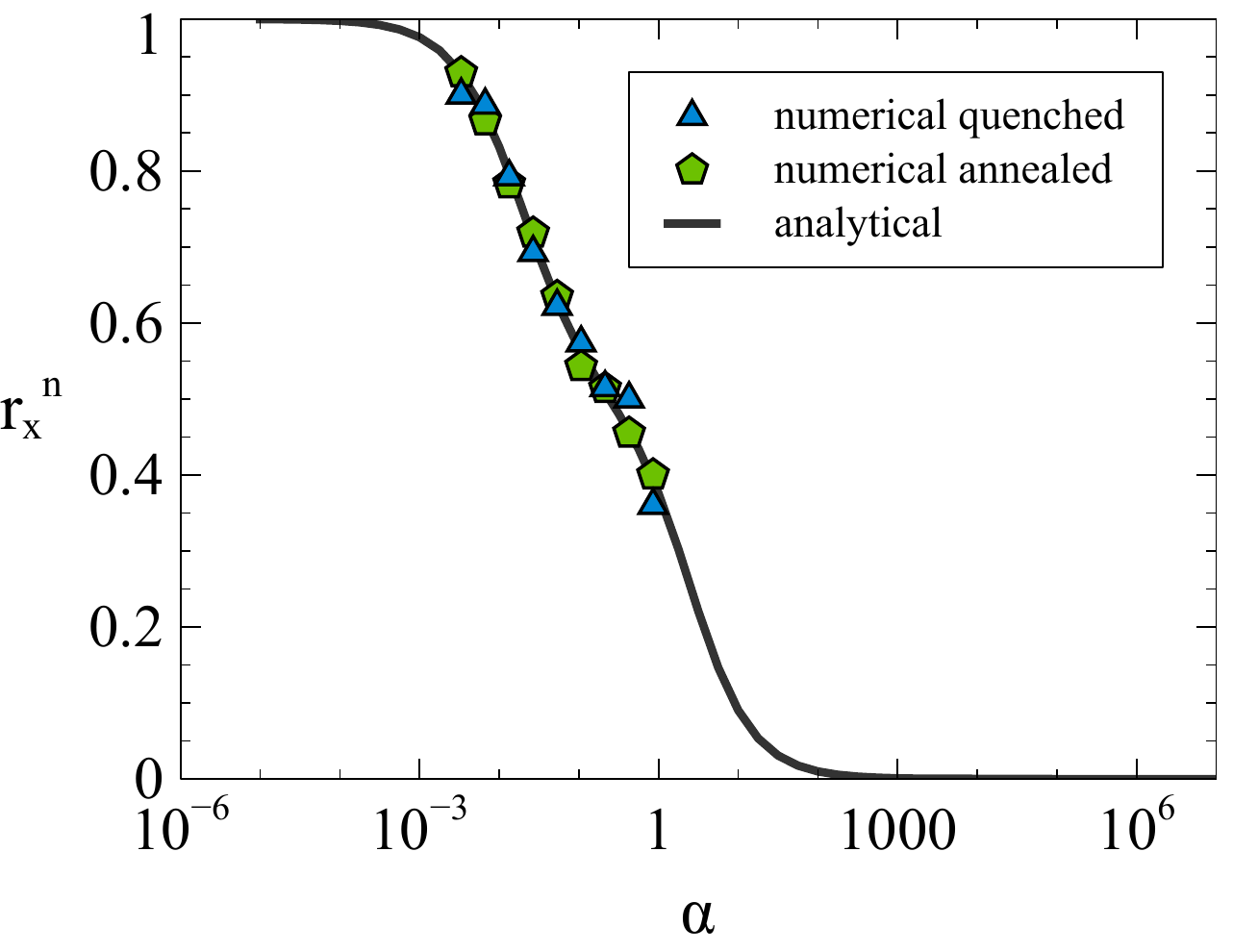}
\caption{Numerical simulations corresponding to annealed and quenched average are in good agreement with the analytical annealed result. With the increase of the fraction of samples $\alpha$ the reconstruction accuracy improves from random guessing ($r_x^n=1$) to a perfect reconstruction ($r_x^n=0$). Annealed and quenched simulation results are in good agreement, indicating the annealed calculation to be a good approximation to the quenched calculation in this case. The parameters are chosen as: $M=1 - 500$ at $N=500$, $n=2$, $\bar{p}=1/3$ and $\sigma_p=0.05$ for all cell types equally, $\bar{x}=3$ and $\sigma_x=1$ also for all cell types equally, $\tilde{\sigma}_\xi=0.1$.}
\label{fig.AnalyticNumericComparison}
\end{center}
\end{figure}

Solving the saddle-point equations  (\ref{eqn:saddlepointeqns}) numerically for the first non-trivial case of $n=2$ cell types, we are particularly interested in the difference between target and the reconstructed mixtures and concentrations and how this difference depends on the number of samples $M$. To this end, we define a normalized order parameter $r_x^n \equiv \frac{r_x}{2\sigma_x}$, which is zero for perfect reconstruction and one for random guessing from the prior distribution of the target solution. Figure \ref{fig.AnalyticNumericComparison} shows how the reconstruction improves with increasing number of samples $M$. This is not surprising as each additional sample brings different mixing proportions and induces correlations in the measurements $X^\mu_i$ across genes, from which both concentrations in the cell types and mixing proportions can be reconstructed. There is no finite threshold in the number of samples below which reconstruction is impossible, at any non-zero 
value of $\alpha=M/N$ the reconstruction is better than a random choice of concentrations and mixtures. 

To compare these results with numerical simulations, we evaluated the quenched average by drawing target mixtures and concentrations from the Gaussian prior distribution and the measurements $X^\mu_i$ generated from~\eqref{eq:likelihood}. Then Markov chain Monte Carlo (MCMC) sampling is used to draw the reconstructed mixtures and concentrations from the Boltzmann posterior distribution~\eqref{eq.Hprior}. We also simulated the annealed average by including prior for the target mixtures/concentrations into the Hamiltonian and sampling over the target mixtures/concentrations as well. Very good agreement between these numerical and the analytical results is seen in Fig. \ref{fig.AnalyticNumericComparison}.

\section{A sample deconvolution algorithm}

The Boltzmann posterior distribution (\ref{eq.Posterior}) with the Hamiltonian \eqref{eq.Hprior} suggests itself as the basis for a simple sample deconvolution algorithm 
based on a particular set of measurements $X^\mu_i$. Using MCMC sampling of this posterior~\cite{d.Gamerman2006}, the entire space of reconstructions can be explored weighted with the posterior probability. An arbitrary starting configuration ${\mathcal C}_0=\{\mathbf{p},\mathbf{x}\}$ is chosen and from this starting point, a new neighbouring configuration ${\mathcal C}_1$ is generated by randomly increasing or decreasing one of the free parameters by a small amount within the positivity constraints and the constraint on the mixing proportions. The new configuration is accepted with the probability $p_{acc} = \min\left(1, e^{-(\mathcal{H}_1-\mathcal{H}_0)}\right)$ with energies $\mathcal{H}_0$/$\mathcal{H}_1$ corresponding to configuration $\mathcal{C}_0$ and $\mathcal{C}_1$, respectively (Metropolis rule). Since only the ratio of the posterior probability of two configurations enters, the marginal likelihood $P\left( \mathbf{X} \right )$ does not enter the algorithm.

We test this algorithm on artificially created datasets. For this purpose a target solution $\mathbf{x}^T$ and $\mathbf{p}^T$ is generated, drawn randomly from Gaussian distributions with means $\bar{x}_a^T$/$\bar{p}_a^T$ and variances $\sigma_{x,a}^{T 2}$/$\sigma_{p,a}^{T 2}$. Then the measurements $\mathbf{X}^T$ are generated according to eq.~(\ref{eq.samplecoupling1}) by adding Gaussian noise of variance $\sigma_\xi^{T 2}$. The Metropolis rule is used to sample the  posterior (\ref{eq.Posterior}) of the mixtures $p_a^\mu$, the concentrations $x_i^a$, and the parameters describing the prior $\bar{x}_a$/$\bar{p}_a$, $\sigma_{x,a}^2$/$\sigma_{p,a}^2$ and $\sigma_\xi^2$s. In this way only the general shape of the prior distribution needs to be chosen, the actual parameters of the distribution are estimated from the data. 

\begin{figure}
\begin{center}
\includegraphics[width=80.0mm]{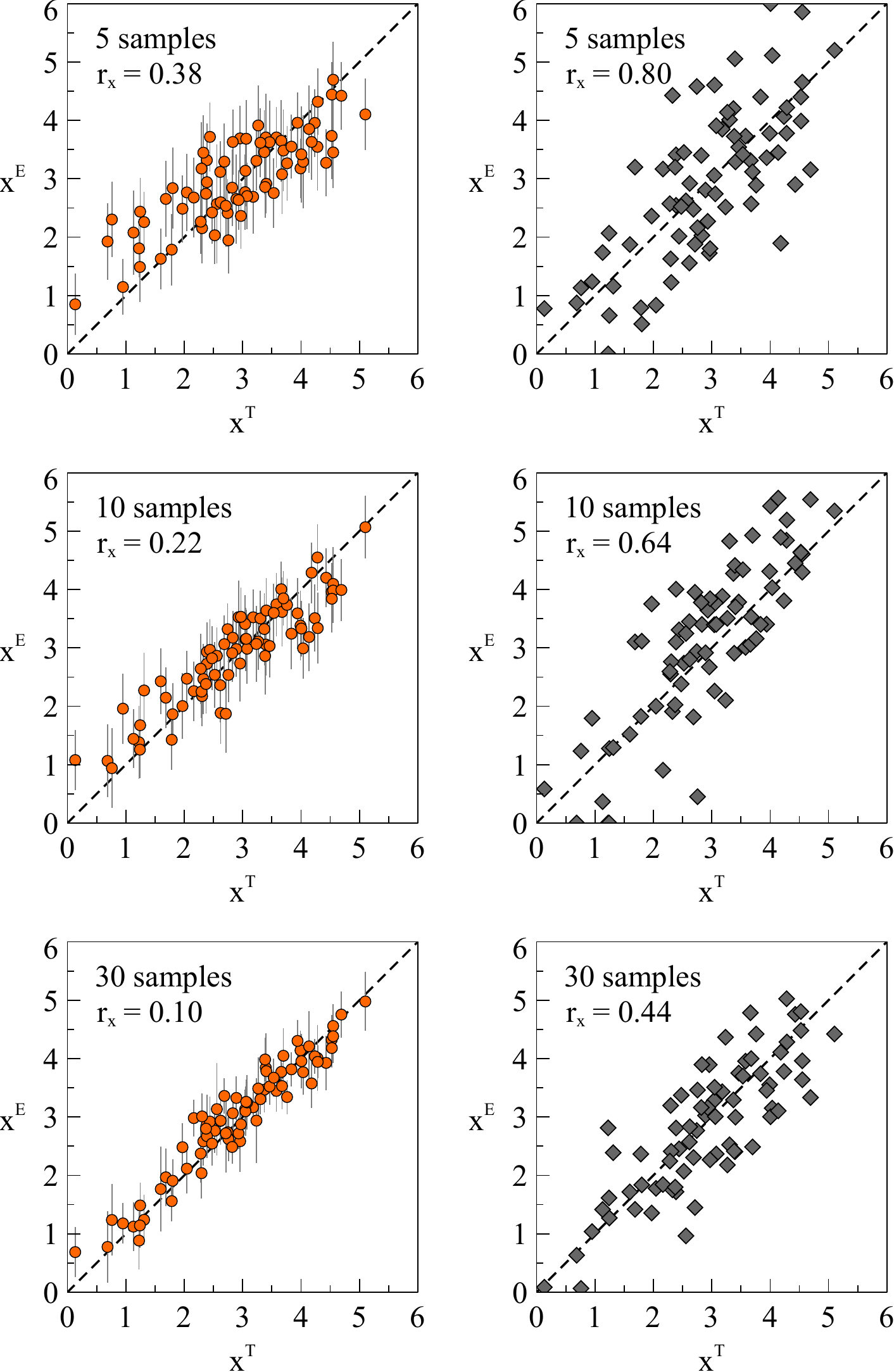}
\caption{Reconstruction by Bayesian algorithm (left plots) and the \textit{deconf} algorithm (right plots). The reconstruction estimates $x^E$ are obtained from the same target solution $\mathbf{x}^T$ for all plots. The reconstruction with the Bayesian algorithm is clearly more accurate: the reconstruction accuracy $r_x = \frac{1}{nN} \sum_{a,i} \left(x_i^{E a}-x_i^{T a}\right)^2$ of the Bayesian algorithm using $5$ samples is comparable to the accuracy of the \textit{deconf} algorithm using $30$ samples. Additionally, the standard deviation of the posterior distribution serves as a natural measure for uncertainty of the estimate, giving rise to the individual errorbars in the case of the Bayesian algorithm. Without knowing $x^T$ they still provide a good estimate for the accuracy of the reconstruction. Parameters used for both algorithms are $N=500$, $n=3$, $M=5, 10$ and $30$, $\bar{p}^T=1/3$ and $\sigma_p^T=0.05$ for all cell types equally, $\bar{x}^T=3$ and $\sigma_x^T=1$ also for all cell types equally, $\sigma_\xi^T=0.1$. In order to distinguish single points with their corresponding error bars the values of only one in twenty of the $500$ genes are plotted.}
\label{fig.NumericalSampling}
\end{center}
\end{figure}

This MCMC sampling allows to sample the regions in phase space where the posterior probability is high. An estimate of the mixtures and concentrations is provided by the mean values of $x_i^a$ and $p_a^\mu$ obtained by averaging over a large number of configurations visited during the sampling process. In addition to this posterior mean, we also compute the standard deviations of $x_i^a$ and $p_a^\mu$ under the posterior. These deviations quantify the remaining uncertainty in the reconstruction, given the limited amount of data and the noise on the data. They can serve as error estimates of the reconstruction when the target solution is unknown. In Fig.~\ref{fig.NumericalSampling} we show for each variable $x_i^a$ the mean and standard deviation (as error bars) under the posterior (\ref{eq.Posterior}) against the targets $x^{T a}_i$. 

The results of this Bayesian sample deconvolution can be compared with those of a different class of sample deconvolution algorithms based non-negative matrix factorization (NMF) to invert the relationship $\mathbf{X}=\mathbf{p} \cdot \mathbf{x}$~\cite{b.Repsilber2010,b.Lahdesmaki2005,b.Venet2001}. Starting with an initial guess of $\mathbf{x}$, the matrix $\mathbf{p}$  is calculated that minimizes the $l_2$-norm $\lVert\mathbf{X}-\mathbf{p}\cdot \mathbf{x}\rVert_\text{F} \equiv \sqrt{\sum\limits_{\mu,i}\left(X^\mu_i - \sum_a p_a^\mu x_i^a\right)^2}$ (Frobenius norm). The minimization proceeds under the additional constraint of non-negative matrix entries. From this estimate of $\mathbf{p}$ an improved estimate for $\mathbf{x}$ is obtained by applying the procedure in turn to $\mathbf{x}$. Iterating these steps  the distance $\lVert\mathbf{X}-\mathbf{p}\mathbf{x}\rVert_\text{F}$ of the reconstructed solution from the data matrix never increases. This corresponds to a minimization of $\mathcal{H}_L$ in eq. (\ref{eq.Hlikelihood}). Convergence of the iterative scheme is thus guaranteed and reaches a local minimum of the Hamiltonian~(\ref{eq.Hlikelihood}).

As an example from the class of NMF algorithms we applied the \textit{deconf} algorithm \cite{b.Repsilber2010} to the same artificial data as used for the Bayesian algorithm. Fig.~\ref{fig.NumericalSampling} shows how the \textit{deconf} is outperformed by the Bayesian approach. $30$ samples are needed by \textit{deconf} to reach a similar reconstruction accuracy as the Bayesian algorithm using only $5$ samples. Another NMF algorithm we tried \cite{b.Venet2001} achieved an even lower accuracy.

\begin{figure}
\begin{center}
\includegraphics[width=50.0mm]{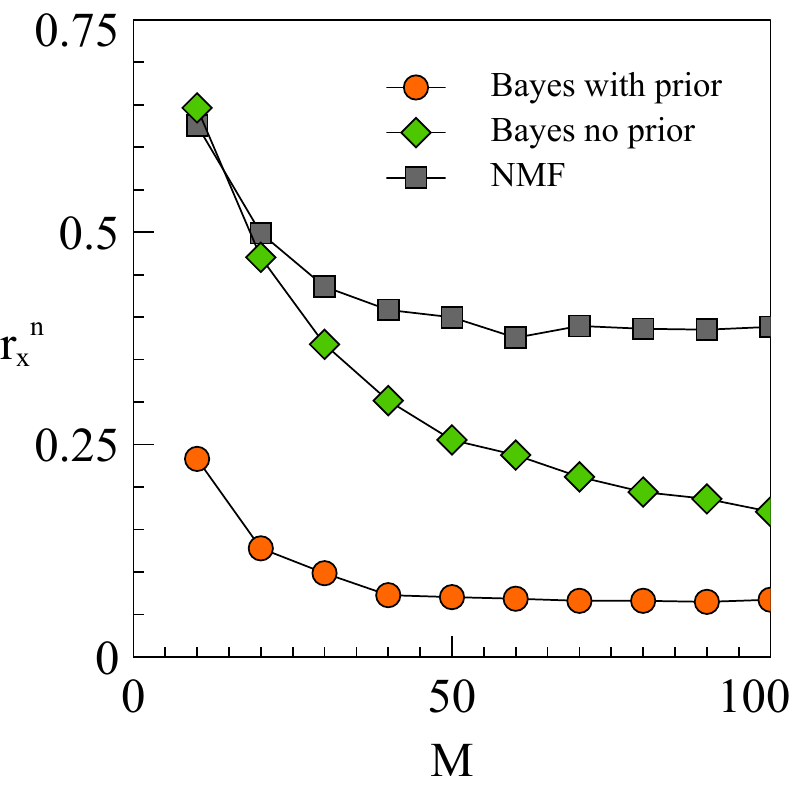}
\caption{The prior information is most important when few samples are available. Then the Bayesian algorithm with prior (orange) clearly outperforms the NMF (gray) but also the Bayesian algorithm without prior (green). Without the use of prior information the Bayes algorithm is performing equally to the NMF algorithm for a small number of samples. With increasing sample number, the Bayes algorithm without prior learns much faster from the additional information, gradually approaching the performance of the Bayes algorithm with prior. Simulation parameters as in fig.~\ref{fig.NumericalSampling}.}
\label{fig.Algorithms_MVar}
\end{center}
\end{figure}

Of course prior information in~\eqref{eq.Hprior} on the distribution of targets facilitates the reconstruction for the Bayesian algorithm. However, the prior is not responsible for the entire difference in performance between the Bayesian and the NMF approach. Fig.~\ref{fig.Algorithms_MVar} shows that the effect of the prior is highest when the number of samples is small. This is to be expected, since the relative contribution of the prior to~\eqref{eq.Hprior} increases when the amount of information coming from the measurements decreases. Even without using prior information, the Bayesian algorithm performs as well as the NMF algorithm in the low sample number regime. For an increasing number of samples the performance of sampling without prior approaches the performance with use of prior information. This is expected as well, since then the relative contribution of the prior to~\eqref{eq.Hprior} becomes asymptotically negligible.
NMF-approaches, formulated as optimization problems, give a point estimate in phase space that reproduces the matrix of measurements $\mathbf{X}$ as closely as possible leading to the well-known problem of overfitting~\cite{m.Opper2003,b.Engel2001}. This is contrasted by the posterior mean estimate of the Bayesian approach, covering the entire phase space weighted by the posterior and leading to a more robust estimate.

For the calculation of the partition function (\ref{eq.PartitionAverage}) we assumed the concrete case of Gaussian distributed residuals. 
For the algorithm, there is no loss of generality involved, any well-behaved probability density can be used in \eqref{eq.Posterior}, leading generally to a Hamiltonian 
that is not quadratic. For the analytic calculation, Taylor expanding such a Hamiltonian around $\mathbf{X}$ would give $\mathcal{H}\left(\mathbf{p},\mathbf{x};\mathbf{X}\right) = a_0 + a_1 (\mathbf{X} - \mathbf{p} \mathbf{x}) + a_2 (\mathbf{X} - \mathbf{p} \mathbf{x})^2 + \cdots $. Our analytical calculation focusses exclusively on the second-order term. 
The first order term alone (apart from not being normalizable), would induce no correlations between the targets $\mathbf{p}^T,\mathbf{x}^T$ and the estimated solution $\mathbf{p},\mathbf{x}$. It may be possible, at least in principle, to evaluate the integrals arising from even polynomials beyond the 
second order, but the resulting expressions will not admit simple order parameters. The Gaussian distributed residuals~\eqref{eq:likelihood} thus define
the non-trivial yet tractable model of sample deconvolution.  

\section{Outlook}
In summary, we have developed a Bayesian model for reconstructing cell-type specific gene concentrations from samples containing an unknown mixture of cell types with unknown concentrations. Formulating the problem in the language of statistical mechanics, we have obtained an analytical solution for the reconstruction accuracy using the annealed approximation for the specific case of $n=2$ cell types. This solution can be extended easily for any finite number of cell types. Our model also suggest a simple algorithm based on the MCMC sampling of the solution space weighted by the posterior distribution of the Bayesian model. This turns out to outperform methods based on minimizing the distance between the matrix of measurements $\mathbf{X}$ and the matrix product of mixing proportions $\mathbf{p}$ and concentrations $\mathbf{x}$. 

As ever, the proper choice of the prior may be a delicate step, and Gaussian priors need not optimally describe real datasets. A study based on experimental datasets would be needed to settle this issue. But if a better choice for the prior is found, it would be straightforward to implement into the algorithm, leaving the rest of the Bayesian framework unchanged.

{\bf Acknowledgement:} This study was funded by the BMBF through Sybacol. We gratefully acknowledge discussions with Roman M\"uller, Bernhard Schermer, and Thomas Benzing from the Kidney Research Center Cologne.

\end{document}